\documentstyle[12pt]{article}
\begin{document} 
\vbox {\vspace{6mm}}
 
\begin{center}
{\large \bf TIME-DEPENDENT INVARIANTS FOR DIRAC EQUATION AND 
NEWTON--WIGNER POSITION OPERATOR}\\[7mm]
{V.I. Man'ko\footnote{on leave from  Lebedev Physical Institute, Moscow, 
Russia} and R.V. Mendes}\\
{\it Grupo de F\'isica--Matem\'atica, Complexo II, Universidade de Lisboa\\ 
Av. Prof. Gama Pinto, 2, 1699 Lisboa Codex, Portugal}\\[5mm]
\end{center}
 
\vspace{2mm}

\begin{abstract}
For Dirac equation, operator-invariants containing explicit 
time-dependence in parallel to known time-dependent invariants of 
nonrelativistic Schr\"odinger equation are introduced and discussed. 
As an example, a free Dirac particle is considered and new invariants 
are constructed for it. The integral of motion, which is initial
Newton--Wigner position operator, is obtained explicitly for a free
Dirac particle. For such particle with kick modeled by delta-function
of time, the time-depending integral, which has physical meaning of
initial momentum, is found.
\end{abstract}

\section{Introduction}

\noindent

The usual time-independent integrals of motion of stationary quantum 
system like energy or angular momentum are related to symmetry of the
Hamiltonian, which is connected with time translational symmetry or
rotational symmetry of the potential~\cite{Messiah}. For nonrelativistic
quantum systems with nonstationary Hamiltonians (for example, for the
parametric oscillator), quadratic in position and momentum integrals
of motion of classical problem, known long time ago~\cite{Er1880},
were shown to exist for quantum problem as well~\cite{Lewis}.
A connection of finding wave function of nonstationary Hamiltonians
with existence of time-dependent integrals of motion has been also 
elaborated~\cite{LR}.

For the same problem of nonrelativistic parametric oscillator, linear 
in position and momentum time-dependent integrals of motion have been
found~[5--7]
and have been related to the Green 
function of Schr\"odinger equation~\cite{dod75}. A connection of the 
integrals of motion with quantum propagator turns out to be general 
property related to Schwinger action principle~\cite{urru}. Futhermore, 
an extension of the N\"other theorem from time-independent integrals of 
motion to time-dependent integrals of motion has been 
done~[10--13].
The method of finding 
time-dependent invariants was developed in recent 
works~[14--17].

The time-independent integrals of motion as energy and angular momentum
are used as well in relativistic problems of scalar and spinor particles
in the framework of Klein--Gordon or Dirac equations~\cite{Messiah}. 
But up to our knowledge till now the time-dependent integrals of motion 
for relativistic quantum systems have not been discussed and there is 
still no one example of time-dependent integral of motion for 
relativistic quantum system.

The aim of this work is to find new time-dependent integrals of motion
for free Dirac particle and to extend the results known for 
nonrelativistic propagator of Schr\"odinger equation, which may be 
obtained as eigenfunction of time-dependent invariants~\cite{dod75}
to the case of Dirac equation.

The problem of position operator in relatvistic domain is more 
complicated than for Schr\"odinger wave mechanics. There are known 
results for Dirac equation to introduce an operator playing role of 
position operator~\cite{Newton,Wigner}. Newton--Wigner position operator
is intensively discussed, in recent publications as well~\cite{SilPr}.
Thus, another goal of our work is to construct a time-dependent
integral of motion, which coincides with Newton--Wigner position
operator for initial time moment. Production of electron--positron 
pairs in heavy ion collisions~\cite{schweppe} was studied recently in 
frame of the model of short pulses described by delta-kickes~\cite{Mendes}.
Dirac equation was used in which a potential was switched in for
very short time corresponding to ultrafast interaction of high energy 
of colliding ions. In this connection, it is worthy to investigate
what type of integrals of motion does exist in such potential. We
construct explicitly expressions for some time-dependent integrals
of motion for Dirac equation with short kickes modeled by delta-function
of time. 

The paper is organized as follows. In Section~2, we review the known 
time-dependent invariants for free nonrelativistic scalar Schr\"odinger
particle (readers having interest to detail review we address to 
Refs.~\cite{Vol183,mal79}\,) and generalize the nonrelativistic 
invariants to the case of free Dirac particle. In Section~3, we discuss
the time-dependent invariants and their properties for Dirac particle
in electromagnetic field. In Section~4, we discuss Newton--Wigner 
position operator and its time-dependent analog, which is 
an integral of motion for Dirac free particle. In Section~5, an 
extension of the integrals of motion found for free Dirac particle
is done for the kicked Dirac particle in frame of the delta-pulse 
model.

\section{Invariants for Free Dirac Particle}
 
\noindent

Let us consider a free Dirac equation in standard form
\begin{equation}\label{1}
i\,\frac {\partial \psi }{\partial t}=
(\alpha \hat p+\beta m)\psi \,;\qquad \hbar =1=c\,, 
\end{equation}
where $p_i=-i\,\partial /\partial x_i~~
(i=1,\,2,\,3)$ are the components of 3--vector $\hat p,\,m$ is constant 
mass, and the 4$\times $4--matrices 
$\alpha _i$ and $\beta $ have the form
\begin{equation}\label{2}
\alpha =\left( \begin{array}{clcr}
0&\sigma \\
\sigma &0\end{array}\right);~~~~~~~~~~~~~~~~~~
\beta =\left( \begin{array}{clcr}
1&0\\
0&-1\end{array}\right),
\end{equation}
where the vector-matrix $\alpha $ has three components
$\alpha _i.$ The Pauli 2$\times $2--matrices which are components
of vector-matrix $\sigma $ are chosen to be
\begin{equation}\label{3}
\sigma _1=\left( \begin{array}{clcr}
0&1\\
1&0\end{array}\right);~~~~~~~~~~~~~~~~
\sigma _2=\left( \begin{array}{clcr}
0&-i\\
i&0\end{array}\right);~~~~~~~~~~~~~~~~
\sigma _3=\left( \begin{array}{clcr}
1&0\\
0&-1\end{array}\right).
\end{equation}
The aim of this section is to discuss for a free Dirac particle the 
existence of a time-dependent integral of the motion.

Let us remind that for a free Schr\"odinger particle with the 
Hamiltonian
\begin{equation}\label{4}
H=\frac {\hat p^2}{2m}\,;~~~~~~~~~~p_i=
-i\,\frac {\partial }{\partial x_i}\,;~~~~~
i=1,\,2,\,3\,,
\end{equation}
there exist the following invariants: momentum operator
described by 3--vector
\begin{equation}\label{5}
\hat p_0(t)=\hat p\,,
\end{equation}
which commutes with the Hamiltonian (\ref{4}), and the initial
coordinate vector-operator $\hat x_0(t)$ with the
components 
\begin{equation}\label{6}
x_{0i}(t)=x_i+
\frac {it}{m}\,\frac {\partial }{\partial x_i}\,.
\end{equation}
Here $x_i$ are components of 3--vector $\hat x.$
The vector-operators $\hat x_0(t)$ and $\hat p_0(t)$
have the property that
\begin{equation}\label{7}
\langle \psi (t)\mid \hat x_0(t)\mid \psi (t)\rangle
=\langle \psi (0)\mid \hat x\mid \psi (0)\rangle
\end{equation}
and
\begin{equation}\label{8}
\langle \psi (t)\mid \hat p_0(t)\mid \psi (t)\rangle =\langle 
\psi (0)\mid \hat p\mid \psi (0)\rangle
\end{equation}
for any solution to the Schr\"odinger equation. At the same time,
these operators may be expressed in terms of the evolution
operator $\hat U(t)$ of the free Schr\"odinger particle.
This evolution operator satisfies the equation
\begin{equation}\label{9}
i\,\frac {\partial \hat U}{\partial t}=
\frac {\hat p^2}{2m}\,\hat U
\end{equation}
with the initial conditions 
$$
\hat U(0)=\hat E\,.
$$
The operator $\hat E$ is the identity operator. In terms of the
operator $\hat U(t),$ the invariant $\hat x_0(t)$ (\ref{6}) is 
expressed as follows 
\begin{equation}\label{10}
\hat x_0(t)=\hat U(t)\,\hat x\,\hat U^{-1}(t)\,,
\end{equation}  
and the operator $\hat p_0(t)$ (\ref{5}) is expressed in the
similar manner 
\begin{equation}\label{11}
\hat p_0(t)=\hat U(t)\,\hat p\,\hat U^{-1}(t)\,.
\end{equation}  
The operator $\hat U(t)$ is the solution to Eq. (\ref{9}) and has the 
form
\begin{equation}\label{12}
\hat U(t)=\exp \left (-it\,\frac {\hat p^2}{2m}\right )\,.
\end{equation}
Thus, the exact forms of the operators $\hat x_0$
and $\hat p_0$ are the result of commuting of the 
operator $\hat U$ (\ref{12}) either with operator $\hat p$ or with
operator $\hat x$ in formulas (\ref{11}) and (\ref{10}),
correspondingly. At the same time, these operators satisfy
the system of equations
\begin{eqnarray}
\frac {\partial \hat x_0}{\partial t}+i\,[H,\hat x_0]&=&0\,;\label{13}\\
\frac {\partial \hat p_0}{\partial t}+i\,[H,\hat p_0]&=&0\,.\label{14}
\end{eqnarray}
For this simple problem, we can use a trick to guess 
the form of the operators~(\ref{5}) and (\ref{6}) on the basis of
classical intuition. 

Let us write down the classical solution for the free motion equation. 
We have
\begin{eqnarray}\label{15}
p\,(t)&=&p_0\,;\nonumber\\
x\,(t)&=&x_0+v_0t\,;\\
p&=&mv\,.\nonumber
\end{eqnarray}
Here $p_0$ and $v_0(p_0)$
are the initial values of the coordinate and velocity
(momentum) of the particle. Expressing   
these initial values in terms of the instant values
$p\,(t)$ and $x\,(t),$ in view of~(\ref{15}), we have
\begin{eqnarray}
p_0&=&p\,(t)\,;\label{16}\\
x_0&=&x-t\,\frac {p\,(t)}{m}\,.\label{17}
\end{eqnarray}
``Quantizing'' these initial values we obtain the quantum 
invariants~(\ref{5}) and (\ref{6}). 

From formulas~(\ref{10}) and 
(\ref{11}), it follows that the propagator of a free Schr\"odinger 
particle $G\,(x,x^\prime ,t)$ satisfies the system of equations
\begin{eqnarray}
\hat x_0(t)\,G\,(x,x^\prime ,t)&=&x^\prime G\,(x,x^\prime ,t)\,;
\label{18}\\
\hat p_0(t)\,G\,(x,x^\prime ,t)&=&
i\,\frac {\partial }{\partial x^\prime }G\,(x,x^\prime ,t)\,.
\label{19}
\end{eqnarray}
Thus, knowing invariants (\ref{5}) and (\ref{6}) we have the 
equations for matrix elements of evolution operator in the coordinate 
representation.

Now let us discuss what properties of the Schr\"odinger
invariants $\hat p_0$ and $\hat x_0$ may be generalized
for the Dirac particle. It is the general problem. But first we 
have to consider the problem applying it to the simplest case of 
a free particle. Equation (\ref{1}) is the equation of the same 
evolutionary type as the Schr\"odinger equation. Thus, the 
4$\times $4--matrix  $U_D(t)$ may be introduced
as evolution operator. This operator-matrix satisfies the
equation
\begin{equation}\label{20}
i\,\frac {\partial U_D}{\partial t}=H_DU_{D}\,,
\end{equation}
where the Hamiltonian $H_D$ is the 4$\times $4--operator-matrix
\begin{equation}\label{21}
H_D=\alpha \hat p+\beta m\,.
\end{equation}
The solution to (\ref{20}) has the same form that in the Schr\"odinger
case
\begin{equation}\label{22}
U_D(t)=\exp \,(-iH_Dt)\,.
\end{equation}
Let us now generalize formulas~(\ref{10}) and (\ref{11}) of the 
Schr\"odinger case to the Dirac case. We have the operators
\begin{eqnarray}
p_{0D}(t)&=&\exp \,(-iH_Dt)\,\hat p\,\exp \,(iH_Dt)\,;\label{23}\\
x_{0D}(t)&=&\exp \,(-iH_Dt)\,\hat x\,\exp \,(iH_Dt)\,.\label{24}
\end{eqnarray}
Equation (\ref{23}) immediately gives the equality for 
$p_{0D}(t),$ namely,
\begin{equation}\label{25}
p_{0D}(t)=\hat p\,,
\end{equation}
as well as it was in the Schr\"odinger case, since the 
Hamiltonian~(\ref{21}) of the Dirac free particle commutes with 
the momentum $\hat p.$ In order to calculate expression~(\ref{24}), 
we could use the Campbell--Hausdorf formula
\begin{equation}\label{26}
e^BAe^{-B}=A+[B,A]+\frac {1}{2!}\,[B,[B,A]]+
\cdots +\frac {1}{n!}\,[B,[B[\ldots [B,A]]]\ldots ]+\cdots \,.
\end{equation}
But since the first commutator (if one takes $B=-itH_{D}$
and $A=x_{i}$ which is the i-th component of the vector 
$\hat x\,)$
\begin{equation}\label{27}
[B,A]=[-it\,(\alpha _kp_k+\beta m),\,x_i]=
-it\,(-i\delta _{ik})\alpha _k=-t\alpha _i
\end{equation}
is not $c$--number, we need to calculate the next commutators 
in~(\ref{26})
\begin{equation}\label{28}
[B,[B,A]]=[-it\,(\alpha _{k}p_{k}+\beta m),-t\alpha _{i}]=
it^2(\,[\alpha _{k}\alpha _{i}]p_{k}+m[\beta \alpha _{i}]\,)\,.
\end{equation}
The matrices $\alpha _i$ and $\beta $ obey the known algebra
\begin{eqnarray}\label{29}
\alpha _i\alpha _k+\alpha _k\alpha _i&=&2\,\delta _{ik}\,;\nonumber\\
\alpha _i\beta +\beta \alpha _i&=&0\,;\\
\alpha _i^2=\beta _i^2&=&1\,.\nonumber
\end{eqnarray}
From this algebra, taking into account~(\ref{28}) it is not obvious 
how to use these formulas for calculating each term in series~(\ref{26}) 
as well as the total sum.

To evaluate this sum, we use the specific property of the Dirac 
Hamiltonian for free particle. Let us introduce $-it=\tau .$ 
Then the matrix $H_D$ has the form 
\begin{equation}\label{30}
H_{D}=\left( \begin{array}{clcr}
m&\sigma \hat p\\
\sigma \hat p&-m\end{array}\right ).
\end{equation}
The square of this matrix is the following diagonal matrix
\begin{equation}
H_D^2=\left( \begin{array}{clcr}
m^2+\hat p^2&~~~~0\\
0&m^2+\hat p^2\end{array}\right )=\varepsilon ^2E\,,
\end{equation}
where
\begin{equation}\label{31}
\varepsilon ^2=\hat p^2+m^2
\end{equation}
and $E$ is the 4$\times $4--identity matrix. Then one has
\begin{equation}\label{32}
H_D^3=\varepsilon ^2H_{D}\,,~~H_D^4=\varepsilon ^4E\,,\ldots \,,\,
H_D^{2n-1}=\varepsilon ^{2n-1}\frac {H_D}{\varepsilon }\,,~~
H_D^{2n}=\varepsilon ^{2n}E\,.
\end{equation}
Thus,
\begin{equation}\label{33}
e^{\tau H_D}=E+\tau \varepsilon H_D+
\frac {\varepsilon ^2\tau ^2}{2!}E+
\frac {\varepsilon ^3\tau ^3}{3!}\frac {H_D}{\varepsilon }+\cdots
=E\cosh \varepsilon \tau +\frac {H_D}{\varepsilon }
\sinh \varepsilon \tau \,.
\end{equation}
This operator in the matrix form is
\begin{equation}\label{34}
e^{\tau H_D}=\left( \begin{array}{clcr}
\cosh \varepsilon \tau +\varepsilon ^{-1}m\sinh \varepsilon \tau &
~~~~\varepsilon ^{-1}\sigma \hat p\sinh \varepsilon \tau \\
\varepsilon ^{-1}\sigma \hat p\sinh \varepsilon \tau &
\cosh \varepsilon \tau -
\varepsilon ^{-1}m\sinh \epsilon \tau \end{array}\right ).
\end{equation}
The matrix $\varepsilon ^{-\tau H_D}$ differs only by sign
before the term with sinus hyperbolic. At $\tau =0\,,$ this 
matrix coincides with the identity matrix $E\,.$ 

The Green function $G_{ik}(x,x^\prime ,t)$ is 
the 4$\times $4--matrix $(i,\,k=1,\,2,\,3,\,4)$ containing 
matrix elements of the evolution operator $U_{D}(t)\,.$ It 
is connected by the relation similar to~(\ref{18}) and (\ref{19}) 
with constructed invariants $x_{0D}(t)$ and 
$p_{0D}(t)=\hat p$ which are 4$\times $4-matrices as well. So
\begin{equation}\label{35}
p_{0D}(t)\,G_{ik}(x,x^\prime ,t)=
i\,\frac {\partial }{\partial x^\prime }\,G_{ik}(x,x^\prime ,t)
\end{equation}
and
\begin{equation}\label{36}
(x_{0D}(t)\,)_{ii^\prime }\,G_{i^\prime k}(x,x^\prime ,t)
=x_{i}^{\prime }G_{ik}(x,x^\prime ,t)\,.
\end{equation}
Here we have in mind summation over two equal indices $i^\prime $
from 1 to 4 and no summation over index $i\,.$ The operators in 
the matrices $x_{0D}$ and $p_{0D}$ act on the first argument $x$ 
of the propagator $G_{ik}(x,x^\prime ,t).$ 

Now we will write down four matrix elements of the matrix 
$x_{0D}(t).$ This matrix consists of four 2$\times $2--blocks. 
We will label these blocks as 
$(x_0)_{11},\,(x_0)_{12},\,(x_0)_{21},$
and $(x_0)_{22}\,.$ (In the case of the free Dirac particle, the 
matrix structure of the operator $p_{0D}(t)$ is simple  because
it is proportianal to unity 4$\times $4--matrix.) Thus, the 
2$\times $2--block $(x_0(t)\,)_{11}$ of invariant $x_{0D}$ is the 
following one 
\begin{eqnarray}\label{37}
(x_{0}(t)\,)_{11}&=&\cosh \varepsilon \tau \,\hat x\,
\cosh \varepsilon \tau 
-\frac {m^2}{\varepsilon }\sinh \varepsilon \tau \,\hat x\,
\frac {1}{\varepsilon }\sinh \varepsilon \tau \nonumber\\
&+&m\left [\frac {1}{\varepsilon }\sinh \varepsilon \tau \,
\hat x\,\cosh \varepsilon \tau -\cosh \varepsilon \tau \,\hat x\,
\frac {1}{\varepsilon }\sinh \varepsilon \tau \right ]\nonumber\\
&-&\frac {\sigma \hat p}{\varepsilon }\sinh \varepsilon \tau \,
\hat x\,\frac {1}{\varepsilon }\sinh \varepsilon \tau \,.
\end{eqnarray}
Here
\begin{equation}\label{38}
\varepsilon =\sqrt {m^2+\hat p^2}\,;~~~p_i=
-i\,\frac {\partial }{\partial x_i}~~(i=1,\,2,\,3)\,;~~~\tau =-it\,.
\end{equation}
We must take into account the order of the operators in the
product in this formula because they do not commute. The next
block $(x_0(t)\,)_{12}$ has the form
\begin{eqnarray}\label{39}
(x_0(t)\,)_{12}&=&
m\left [\frac {\sigma \hat p}{\varepsilon }\sinh \varepsilon \tau 
\,\hat x\,\frac {1}{\varepsilon }\sinh \varepsilon \tau -
\frac {1}{\varepsilon }\sinh \varepsilon \tau \,\hat x\,
\frac {\sigma \hat p}{\varepsilon }\sinh \varepsilon \tau \right ]
\nonumber\\
&+&\frac {\sigma \hat p}{\varepsilon }\sinh \varepsilon \tau \,\hat x\,
\cosh \varepsilon \tau -\cosh \varepsilon \tau \,\hat x\,
\frac {\sigma \hat p}{\varepsilon }\sinh \varepsilon \tau \,.
\end{eqnarray}
Of course, we can use the commutator $[x_i,\,p_k]=i\delta _{ik}\,,
~(i,\,k=1,\,2,\,3)$ in order to rewrite the terms in these formulas.
For the block $(x_{0})_{21}\,,$ we have the formula  
\begin{eqnarray}\label{40}
(x_{0}(t)\,)_{21}&=&
m\left [\frac {1}{\varepsilon }\sinh \varepsilon \tau \,\hat x\,
\frac {\sigma \hat p }{\varepsilon }\sinh \varepsilon \tau \right ]-
\frac {\sigma \hat p}{\varepsilon }\sinh \varepsilon \tau \,\hat x\,
\frac {1}{\varepsilon }\sinh \varepsilon \tau \nonumber\\
&+&\frac {\sigma \hat p}{\varepsilon }\sinh \varepsilon \tau \,\hat x\,
\cosh \varepsilon \tau -\cosh \varepsilon \tau \,\hat x\,
\frac {\sigma \hat p}{\varepsilon }\sinh \varepsilon \tau \,.
\end{eqnarray}
The block $(x_0)_{22}$ of the 4$\times $4--matrix-invariant
$x_{0}(t)$ has the form
\begin{eqnarray}\label{41}
(x_0(t)\,)_{22}&=&
\cosh \varepsilon \tau \,\hat x \,\cosh \varepsilon \tau -
\frac {m^2}{\varepsilon ^2}\sinh \varepsilon \tau \,\hat x\,
\sinh \varepsilon \tau \nonumber\\
&+&m\left [\frac {1}{\varepsilon }
\cosh \varepsilon \tau \,\hat x\,\sinh \varepsilon \tau 
-\frac {1}{\varepsilon }\sinh \varepsilon \tau \,\hat x\,
\cosh \varepsilon \tau \right ]\nonumber\\
&-&\frac {\sigma \hat p}{\varepsilon }
\sinh \varepsilon \tau \,\hat x\,
\frac {\sigma \hat p}{\varepsilon }
\frac {\sinh \varepsilon \tau }{\varepsilon }\,.
\end{eqnarray}
From the structure of these blocks, it is evident that it is not easy 
to obtain the invariant-operator $\hat x_0$ using the Campbell--Hausdorf
formula. Thus, we have generalized the integrals
of the motion for the free Schr\"odinger particle to the case of the
free Dirac particle. By constructing the 4$\times $4--matrix-operator
$\hat p_{0D}$ and $\hat x_{0D}$ obey the Weyl--Heisenberg algebra
with commutation relations
\begin{equation}\label{42}
[p_{0Dk},\,x_{0Dl}]=-i\delta _{lk}\,.
\end{equation}
From these operators, it is possible to construct some invariants
with different commutation relations. So taking the operator
\begin{equation}\label{43}
L_i(t)=\varepsilon _{ilk}\,x_{0Dl}\,p_{0Dk}\,,
~~~~~~i,\,k,\,l=1,\,2,\,3,
\end{equation}
one can check that the three time-dependent operators obey
the commutation relations of the angular momentum
\begin{equation}\label{44}
[L_{i}(t),L_{k}(t)]=i\,\varepsilon _{ikl}\,L_{l}(t)\,.
\end{equation}
On the other hand, at $t=0$ these operators 
coincide with the orbital angular momentum operator
$L(t=0)=[\hat x$$\times $$\hat p],$ which does not conserve,
but the operator (\ref{43}) is invariant. Along with the
operators $p_{0D},$ the invariants $L(t)$ form the Lie
algebra of the generators of the motion group in three-dimensional 
space. For the free Dirac particle, the
conserved operator (which is the sum of orbital and spin
momenta) $j=[\hat x$$\times $$\hat p]+\sigma /2\,.$ It commutes with
the Dirac energy operator $H_{D}\,.$ Due to this, one can
write down the equality  
\begin{equation}\label{45}
U_D\,L(t=0)\,U_D^{-1}+\frac {1}{2}\,U_D\,\sigma \,U_D^{-1}=
L+\frac {\sigma }{2}\,.
\end{equation}
Thus, the orbital angular momentum described by the operator
$L(t)$ may be expressed as
\begin{equation}\label{46}
L(t)=L(t=0)+\frac {\sigma }{2}-\frac {1}{2}\,U_D\,\sigma \,U_D^{-1}\,.
\end{equation}
The matrix $U_D(t)\,[\sigma /2]\,U_D^{-1}(t)=S(t)$ is the 
following integral of the motion which describes the 
initial spin operator. It may be obtained in the form of
4$\times $4--matrix operator using formulas 
(\ref{24}) and (\ref{37})--(\ref{41})
by the simple substitution: $\hat x\rightarrow \sigma $ in
all these formulas, all other terms in the formulas
being unchanged.

\section{General Consideration of Operator-Invariants 
for Dirac Particle}

\noindent

The analysis of the previous section was based on the fact
that the free Dirac particle motion may be described by 
the equation of an evolutianary type with the concrete 
Hamiltonians (\ref{20})--(\ref{21}). After that, we used the 
free propagator or evolution operator for free Dirac particles 
in order to construct the invariants 
$p_{0D}(t),\,x_{0D}(t),$ and $L(t)$
which contained dependence on time. But it is clear that
the same procedure may be fulfilled not only for the free
Dirac particle. Any evolutianary type equation for bispinor
$\psi $ of the kind
\begin{equation}\label{d1}
i\,\frac {\partial \psi}{\partial t}=H_D\psi 
\end{equation}
with the Hamiltonian
\begin{equation}\label{d2}
H_D=\left \{\alpha \left [\hat p-\frac {e}{c}\,A(x,t)\right ]
+\beta m+V(x,t)\right \}\,,
\end{equation}
in which we have taken into account also the possibility of
presence of electric and magnetic fields, posesses the 
evolution operator or propagator $\hat U_D(t),$ which makes the 
bispinor $\psi (t)$ from the initial bispinor $\psi (0)$ 
satisfying Eq. (\ref{d1})
\begin{equation}\label{d3}
\psi (t)=U_{D}(t)\,\psi (0)\,.
\end{equation}
If for the more complicated cases with electric and magnetic
fields the evolution operator $\hat U_D(t)$ is found, we 
can construct invariants $\hat I_D(t)$ by the formula
\begin{equation}\label{d4}
I_D(t)=U_D(t)\,I_D(0)\,U_D^{-1}(t)\,.
\end{equation}
The properties of the operator $I_D(t)$ are very similar
to the properties of corresponding integrals of the motion
in the case of Schr\"odinger equation and, of course, these  
Dirac invariants-operators tend to the integrals of the
motion of the Schr\"odinger particle in nonrelativistic limit.
Some theorems about Schr\"odinger integrals of the motion 
may be transfered to the case of the Dirac equation. For example,
a function of the Dirac invariant-operator is another Dirac
invariant-operator. This statement means, that if we have a
function $f(I_D(t)\,),$ this function may be always represented 
in the form
\begin{equation}\label{d5}
f(I_D(t)\,)=U_D(t)\,f(I_D(0)\,)\,U_D^{-1}(t)\,.
\end{equation}
But this formula follows from the obvious property
\begin{equation}\label{d6}
f\left (U_D(t)\,I_D(0)\,U_D^{-1}(t)\right )=
U_D(t)\,f(I_D(0)\,)\,U_D^{-1}(t)\,.
\end{equation}
For example, for any power series with respect to the operator
$I_D(t)$ it can be checked and extended for more 
complicated function. From this property, it follows that
a free Dirac particle has also the following invariants: 
$$
p_{0D}^2,\quad x_{0D}^2,\quad (p_{0D}\,x_{0D}),\quad p_{0D}
\,L(t)\,.
$$
The eigenvalues of the Dirac 
invariants-operators do not depend on time. It means 
that if we search for the solution to the equation
\begin{equation}\label{d7}
I_D(t)\,\psi (t)=\lambda (t)\,\psi (t)\,,
\end{equation}
where bispinor $\psi (t)$ is a solution to the Dirac equation,
the function $\lambda (t)$ is constant and does not depend
on time. This statement follows from simple checking. We have
\begin{equation}\label{d8}
U_D(t)\,I_D(0)\,U_{D}^{-1}(t)\,U_{D}(t)\,\psi (0)=\lambda (t)
\,U_D(t)\,\psi (0)\,.
\end{equation}
After simple algebra, it is obvious that $\lambda (t)$ is 
eigenvalue of the time-independent operator $I_{D}(0)$
and due to this it does not depend on time as well.

The third statement  which may be generalized from the
Schr\"odinger example is that if we have a solution to the
Dirac equation $\psi (t)$ and act on it by the Dirac invariant
operator $I_D(t),$ the new bispinor $\varphi (t)$ such that
\begin{equation}\label{d9}
\varphi (t)=I_{D}(t)\,\psi (t)
\end{equation}
turns out to be the solution to the Dirac equation, too. This 
statement may be proved by simple substitution. For the solution
$\psi (t),$ we have the expression
\begin{equation}\label{d10}
\psi (t)=U_D(t)\,\psi (0)\,.
\end{equation}
Then, we have
\begin{equation}\label{d11}
I_D(t)\,\psi (t)=
U_D(t)\,I_D(0)\,U_D^{-1}(t)\,U_D(t)\,\psi (0)\,,  
\end{equation}
but the bispinor
\begin{equation}\label{d12}
\varphi (t)=U_D(t)\,\varphi (0)\,,
\end{equation}
where
\begin{equation}\label{d13}
\varphi (0)=I_D(0)\,\psi (0)\,,
\end{equation}
satisfies the Dirac equation due to the property of the Dirac 
propagator $U_D(t)$. 

The matrix elements of the Dirac evolution 
operator $U_D(t)$ in any representation satisfy the relations
analogous to (\ref{35}) and (\ref{36}) for the free Dirac particle.
In the coordinate representation, for any Dirac Hamiltonian if we 
found the operators $\hat x_0(t)$ and $\hat p_0(t),$
these equations hold.

\section{Newton--Wigner Operator}

\noindent

The integral of motion, which is an initial position operator,
turns out to play important role in considering a nonrelativistic
quantum system~\cite{Vol183,mal79}. Due to this for relativistic
Dirac particle it is worthy to study an analogous integral of
motion, which plays the role of initial position of the particle.
For the Dirac particle, the position operator itself has some
pecularities and it differs from the position operator of 
nonrelativistic particle. The most appropriate operator playing 
the role of the position operator of the Dirac particle was
considered by Newton and Wigner~\cite{Newton,Wigner}.

We will construct the time-dependent integral of motion, which
has the physical meaning of the initial position operator.
It means that this integral of motion at the initial moment of 
time must coincide with the Newton--Wigner position operator.

The Newton--Wigner position operator~\cite{Newton,Wigner} has the 
form (see, for example,~\cite{SilPr}\,)
\begin{equation}\label{nw1}
Q_k=\Lambda _+\left (\hat p\right )\left(1+\gamma _0\right )
\sqrt {\frac {p_0}{p_0+m}}\left (i\,\frac {\partial }{\partial p_k}
\right )\sqrt {\frac {p_0}{p_0+m}}\,\Lambda _+\left (\hat p\right ),
\end{equation}
where 
$$
p_0^2=p_1^2+p_2^2+p_3^2
$$
and the projector
\begin{equation}\label{nw2}
\Lambda _+\left (\hat p\right )=\frac {1}{2}\left (1+\frac {\alpha 
\hat p +\beta m}{p_0}\right ).
\end{equation}
Also we used for the operator of coordinate in momentum 
representation the notation 
\begin{equation}\label{nw3}
\hat x_i=i\,\frac {\partial }{\partial p_k}\,,\qquad i=1,\,2,\,3\,.
\end{equation}

For free Dirac particle, one can construct 3-vector-operator 
$Q_k\left (t\right ),$ which is integral of motion, using the general
construction and invariant found in previous section, i.e.,
\begin{equation}\label{nw4}
Q_k\left (t\right )=U_D\left (t\right )Q_kU_D^{-1}\left (t\right ).
\end{equation}
Since the operators
\begin{equation}\label{nw5}
U_D\left (t\right )p_0U_D^{-1}\left (t\right )=p_0
\end{equation}
and
\begin{equation}\label{nw6}
U_D\left (t\right )\Lambda _+\left (\hat p\right )
U_D^{-1}\left (t\right )=\Lambda _+\left (\hat p\right )
\end{equation}
are integrals of motion for free particle, the integral of motion,
which for initial time $t=0$ coincides with the Newton--Wigner
position operator is
\begin{equation}\label{nw7}
Q_k\left (t\right )=\Lambda _+\left (\hat p\right )\left (1
+\gamma _0(t)\right )\sqrt {\frac {p_0+m}{p_0}}
x_k\left (t\right )\sqrt {\frac {p_0+m}{p_0}}
\Lambda _+\left (\hat p\right )\,.
\end{equation}
Here the integral of motion $x_k\left (t\right )\equiv x_{0Dk}(t)$ 
is given in coordinate representation by~(\ref{37})--(\ref{41}) 
and in momentum representation by the same 
formulas~(\ref{37})--(\ref{41}), 
in which the operator $\hat x$ is replaced by 
$i\,\partial /\partial \hat p\,.$ The integral of motion
\begin{equation}\label{nw8}
\gamma _0\left (t\right )=U_D\left (t\right )\gamma _0
U_D^{-1}\left (t\right )
\end{equation}
in explicit form is
\begin{equation}\label{nw9}
\gamma _0\left (t\right )=\left (\begin{array}{clcr}
\cosh ^2\varepsilon \tau -\varepsilon ^2\left (m^2-\hat p^2\right )
\sinh ^2\varepsilon \tau &
~~~~-2\varepsilon ^{-1}\sigma \hat p\sinh \varepsilon \tau \left (
\varepsilon ^{-1}\sinh \varepsilon \tau +\cosh \varepsilon \tau 
\right )\\
-2\varepsilon ^{-1}\sigma \hat p\sinh \varepsilon \tau \left (
\varepsilon ^{-1}\sinh \varepsilon \tau -\cosh \varepsilon \tau \right )
&~~~~~\varepsilon ^{-2}\left (m^2-\hat p^2\right )
\sinh ^2\varepsilon \tau -\cosh ^2\varepsilon \tau \end{array}\right ).
\end{equation}
Thus, we get 3-vector-operator $\widehat Q_k\left (t\right ),$ which
is the integral of motion of free Dirac particle. This integral of
motion ``remembers'' its initial maximally localized position in space.

\section{Short Pulses for Dirac Partilce}

\noindent

Let us consider the model, in which the interaction term for Dirac
particle is the term $V\left (x,\,t\right )$ in (\ref{d2}), i.e., we 
have
\begin{equation}\label{Dp1}
H_D=\alpha \hat p+\beta m+V\left (x,\,t\right ).
\end{equation}
We study what influence produces very short pulse of the potential.
It is motivated by discussion~\cite{Mendes} of heavy ion collisions 
so fast passing each the other in process of high energy reaction 
that the perturbation of the potential field lasts very short time 
in comparison with the periods of all motions in the system. 
In this case, the model of the potential $V\left (x,\,t\right )$ may 
be described by $\delta $-function of time
\begin{equation}\label{Dp2}
V\left (x,\,t\right )=\kappa V\left (x\right )\delta \left (t-
t^\prime \right ).
\end{equation}
We study explicitly how the propagator for the free Dirac
particle is changed if the potential of the type~(\ref{Dp2}) 
switches in. This problem is interesting not only in the context 
of studying pair production in the process of charge collisions
but also as the simple exactly solvable model of relativistic
particle with nonstationary Dirac Hamiltonian. Then solving 
Eq.~(\ref{d1}) with the Hamiltonian~(\ref{Dp1}), (\ref{Dp2})  
we can find that the propagator of the particle with 
$\delta $-type interaction in time may be factorized
\begin{equation}\label{Dp3}
U_D\left (t_{\rm f},\,t_{\rm {in}}\right )=U_{\rm f}
\left (t_{\rm f},\,t^\prime \right )
\,\exp \left (-i\kappa V\right )U_{\rm f}\left (
t^\prime ,\,t_{\rm {in}}\right ).
\end{equation}
Here $U_{\rm f}\left (t_2,\,t_1\right )$ is the propagator of 
free particle and the influence of $\delta $-pulse at the moment 
$t^\prime $ gives the factor $\exp \left (-i\kappa V\right ).$ 
The free propagator has been calculated in previous sections and 
is described by formula~(\ref{34}). If we introduce the notation
\begin{equation}\label{Dp4}
-i\left (t-t^{'}\right )=\tau _1\,;\qquad -i\left (t^{'}-t_0\right )
=\tau _2\,;\qquad \varepsilon _1=\sqrt {p_1^2+m^2}\,;\qquad
\varepsilon _2=\sqrt {p_2^2+m^2}
\end{equation}
and
\begin{equation}\label{Dp5}
\left (\exp \left [-i\kappa V\right ]\right )_{p_1p_2}=
\int \exp \left [i\left (p_1-p_2\right )x-i\kappa 
V\left (x\right )\right ]\,dx\,,
\end{equation}
in momentum representation, the propagator of the particle 
$G\left (p_1,\,p_2,\,t,\,t_0\right )$ with $\delta $-pulse 
takes the form
\begin{equation}\label{Dp6}
G\left (p_1,\,p_2,\,t,\,t_0\right )=
\left (\exp \left [-i\kappa V\right ]\right )_{p_1p_2}
\,\prod _{i=1}^2
\left( \begin{array}{clcr}
\cosh \varepsilon _i\tau _i
+\varepsilon _i^{-1}\,m\,\sinh \varepsilon _i\tau _i&~~~
\varepsilon _i^{-1}\,\sigma \,p_i\sinh \varepsilon _i\tau _i\\
\varepsilon _i^{-1}\,\sigma \,p_i\sinh \varepsilon _i\tau _i&
\cosh \varepsilon _i\tau _i
-\varepsilon _i^{-1}\,m\,\sinh \varepsilon _i\tau _i
\end{array}\right ).
\end{equation}
That is exactly the form of the Green function for Dirac particle
with $\delta \left (t-t^{'}\right )$-time dependence in the
potential, where $t^{'}$ is the moment of $\delta $-kick. For 
$\kappa =0\,,$ the first term gives the 
$\delta \left (p_1-p_2\right )$-dependence and the product
of two 4$\times $4-matreces in~(\ref{Dp6}) gives one matrix of 
the form~(\ref{34}). 

Let us find out the change of the form of the integral of motion
for free Dirac particle related to the $\delta$-kick, for which
$t^\prime =0.$ Obviously,
for negative times the form of the integrals of motion operators
is the same as for free particle. But for positive times the form
is changed. So, for the form of the invariant, which is initial 
momentum, one obtaines
\begin{equation}\label{Dp7}
\left (\hat p_0\left (t\right )\right )_{pp^{'}}=\Theta \left (-t
\right )p\,\delta \left (p-p^{'}\right )+
\Theta \left (t\right )\int p^{''}\left (\exp \left [i\kappa V
\right ]\right )_{pp^{''}}\left (\exp \left [i\kappa V
\right ]\right )^*_{p^{'}p^{''}}\,dp^{''},
\end{equation}
where
\begin{eqnarray*}
\Theta \left (t\right )&=&1\,,\qquad \qquad t>0\,;\nonumber\\
\Theta \left (t\right )&=&0\,,\qquad \qquad t<0\,.\nonumber
\end{eqnarray*}
The matrix $\left (\exp \left [-i\kappa V\right ]
\right )_{p_1p_2}$ is proportional the unity matrix in four-space.

Thus, we have found the kernel of the time-dependent integral
of motion for the kicked Dirac particle. This integral of
motion plays the role of the initial momentum of the relativistic 
quantum motion.

\section{Conclusion}

\noindent

We have found some time-dependent integrals of motion for
relativistic quantum particles. This consideration of time-dependent
integrals of motion is motivated by the problem of pair creation
in heavy ion collisions. Such processes may be modeled by fast time 
dependence of the interaction potential in Dirac equation. Due to 
this it is interesting to consider the constant of the motion
in such type of models. The results of the paper are given in the 
explicit expressions~(\ref{37})--(\ref{41}) for the analog of the 
initial coordinate operator which is the integral of motion for 
the Dirac particle and in (\ref{Dp7}) for the analog of the initial 
momentum, which is the integral of motion for the kicked Dirac 
particle. We have found the time-dependent integral of motion for 
free Dirac particle given by (\ref{nw7}), which is initial position 
operator by Newton and Wigner. Also the time-dependent integrals 
of motion related to initial spin and angular momentum 
observables~(\ref{46}) of free Dirac particle are the result of 
the analysis presented. It is worthy to study other time-dependent 
integrals of motion, which may be related to the high-energy 
processes modeled by interaction potentials dependent on time.
One should note that in all cases of the relativistic quantum
problems, for which the propagators are known explicitly, it
is possible to find the explicit expressions for the kernels of
time-dependent integrals of motion using the procedure analogous
to the procedure applied for finding the initial momentum
invariant of the kicked Dirac particle.

\section*{Acknowledgments}

\noindent

One of the authors (V.I.M.) would like to acknowledge Grupo de 
F\'isica--Matem\'atica, Complexo II, Universidade de Lisboa
for kind hospitality
and Russsian Foundation for Basic Research under the 
Project No.~96-02-17222 for partial support.

\end{document}